\documentstyle[aps,epsf]{revtex}

\newcommand \boldx {\mbox{\boldmath $x$}}
\newcommand \boldy {\mbox{\boldmath $y$}}
\newcommand \boldh {\mbox{\boldmath $h$}}
\newcommand \boldphi {\mbox{\boldmath $\Phi$}}
\newcommand \sym  {\mbox{SYM}}
\newcommand \as   {\mbox{ASYM}}

\begin{document}
\title{A Replica Approach to Products of Random Matrices}
\author{M. Weigt\thanks{martin.weigt@physik.uni-magdeburg.de} \\
      {\small{\it Institut f\"ur Theoretische Physik,
      Otto-von-Guericke-Universit\"at Magdeburg}}\\
     {\small{\it PSF 4120, 39016 Magdeburg, Germany}}\\[.3cm]
     }
\date{\today }
\maketitle
\begin{abstract}
We analyse products of random $R\times R$ matrices by means of a
variant of the replica trick which was recently introduced for
one--dimensional disordered Ising models. The replicated transfer
matrix can be block--diagonalized with help of irreducible 
representations of the permutation group. We show that the
free energy (or the
Lyapunov exponent) of the product corresponds to the replica
symmetric representation, whereas non--trivial representations
correspond to certain correlation functions.
\end {abstract}
\vskip 1cm

\section{Introduction}

The asymptotic properties of products of random matrices play an 
important role in many physical problems \cite{Luck,Crisanti}. 
In models like disordered one--dimensional magnetic systems they
describe the thermodynamic quantities such as free energy or
correlations, for localization of electronic waves in random potentials 
they are related to the transport properties, see also \cite{Pendry}.
Such products also appear in the context of chaotic dynamical
systems characterizing the divergence of neighboring trajectories.

Although there are many known results on products of random matrices,
some of them even mathematically rigorous, we want to present a 
general replica transfer matrix method. Replicas are known to be
a very powerful but nevertheless somewhat mysterious tool in the statistical
mechanics of disordered systems and related problems \cite{MPV}. In the
case of mean--field models replicas predict the concept of replica
symmetry breaking which is related to a highly nontrivial
ultrametric structure of states in the low temperature phase.

The existence of replica symmetry breaking in low dimensional
systems is not yet clear, see e.g. the argumentation of 
\cite{Newman}. In \cite{RFIM1D} a replica approach to one--dimensional
disordered Ising models was presented. Although there does not exist
any phase transition at nonzero temperature, a rich replica structure
could be observed leading to a 'natural' critrion for replica symmetry breaking 
in this special system which is not related to Parisi's replica symmetry 
breaking scheme for mean--field models. This criterion is based on the 
representation structure of the permutation group and could be deduced
to a large extent with rigorous methods. In the present paper we
generalize the approach of \cite{RFIM1D} to infinite products of random
matrices of finite dimension. We mainly use the 
language of statistical mechanics, i.e. the random matrices
are considered as transfer matrices of one--dimensional models with
finite discrete degrees of freedom and random short--range interactions.
We show that the representation theoretic approach to replica symmetry
breaking is quite general and can be formulated without specifying
a particular one--dimensional model.

The outline of the paper is the following. In Sec. II we introduce
the replicated transfer matrix. For the analysis we need several tools
from the representation theory of the symmetric group. These are
presented in the third section. In Sec. IV the replica symmetric
representation space is considered and the free energy is calculated.
The connection between non--trivial representations of the
symmetric group and connected correlation functions is analyzed
in Sec. V. They will provide a natural criterion for replica
symmetry breaking. In the last section we give a summary and outlook.
Several appendices contain longer calculations or proofs.

\section{The replicated transfer matrix}

We consider $N$ $R\times R$ matrices $T_i, i=1,...,n,$ drawn from a 
single probability distribution $P(T)$, where $R$ is any positive integer.
In the case of an one--dimensional model with random Hamiltonian 
$H=\sum_i H_i(s_i,s_{i+1})$ ($s_i$ can take $R$ different
values) and inverse temperature $\beta$ they are given by 
$T_i=(\exp\{H_i(s_i,s_{i+1})\})$.
For a general distribution these matrices do not
commute. Therefore we cannot find a common system of eigenvectors. In
order to calculate self--averaging quantities such as the free energy
we introduce as usual
the n-fold replicated and disorder averaged partition function,
\begin{eqnarray}
  \label{partfunc}
  \ll Z^n \gg & = & \ll (\mbox{tr} \prod_{i=1}^N T_i )^n \gg \nonumber\\
            & = & (\mbox{tr} \ll T^{\otimes n} \gg)^N\,,
\end{eqnarray}
where $\ll\cdot\gg$ denotes the average with respect to $P(T)$ and 
$\otimes$ the Kronecker product of matrices. 
With this relation we are able to replace the product of $N$ random 
$R\times R$ matrices by the $N$-th power of a single $R^n\times R^n$
matrix which can be analyzed using standard transfer matrix techniques:
 we have to find expressions for the eigenvalues of 
 $T_{n} := \ll T^{\otimes n} \gg$ which enable an analytic continuation
in $n$. The free energy is then given by
\begin{eqnarray}
  \label{repltrick}
  f = -\frac{1}{\beta}\ll \ln Z \gg\;=\; -\frac{1}{\beta}\lim_{n\to 0}
      \partial_n \ll Z^n \gg \;,
\end{eqnarray}
which is dominated by the largest eigenvalue of $T_n$ for $n\to 0$.
Several correlation length can be described by smaller eigenvalues of
the same matrix.

To calculate this, some notations will be introduced. The original matrices 
$T_i$ act on a $R$--dimensional vector space $V$. As a basis
we chose any orthonormalized set of $R$ vectors and denote these by
$|s\rangle,\,s=1,...,R$. Consequently $T_n$ is a linear operator
defined on the $n$--fold tensor product $V^{\otimes n}$ of $V$ with
itself which has dimension $R^n$. The orthonormalized
basis-vectors of this space are chosen naturally as
$|s^1\rangle\otimes|s^2\rangle\otimes...\otimes|s^n\rangle$
$=:|s^1s^2...s^n\rangle$
where $s^a\in \{1,...,R\}$ for all $a=1,...,n$. The matrix elements of $T_n$
are then given by
\begin{eqnarray}
  \label{Tn}
  \langle s^1s^2...s^n | T_n | s'^1s'^2...s'^n \rangle 
  & = &\ll \prod_{a=1}^n \langle s^a| T | s'^a\rangle \gg \nonumber\\
  & = & \ll \prod_{a=1}^n T_{s_a,s'_a} \gg
\end{eqnarray}
for any two basis vectors of $V^{\otimes n}$. 

The average over $P(T)$ produces interactions between the replicas. 
Nevertheless the replicas
are completely equivalent, a renumbering does not change the 
matrix $T_n$. This
leads to a symmetry of the transfer matrix under
replica permutations, i.e. to replica
symmetry of $T_n$. The action of any permutations is 
given by the $R^n$ dimensional representation $D$
of the symmetric group ${\cal S}_n$:
\begin{equation}
\label{Drep}
D(\pi) |s^1s^2...s^n\rangle =  |s^{\pi(1)}s^{\pi(2)}...s^{\pi(n)}\rangle,
\;\; \forall \pi \in {\cal S}_n\;,
\end{equation}
whose operator product with $T_n$ commutes,
\begin{equation}
  \label{commute}
  D(\pi)\;T_n = T_n\;D(\pi)\;, \;\;\forall\pi \in {\cal S}_n\;.
\end{equation}
A direct consequence of equation (\ref{commute}) 
is the closure of any eigenspace of $T_n$
under permutations, these eigenspaces define a subrepresentations of $D$ which
in the most general case are irreducible. Further reducibilities would be a 
hint to a further hidden symmetry.

Consider an element $Y$ of the group algebra ${\bf s}_n$ of ${\cal S}_n$, 
i.e. $Y$ is a 
linear combination of permutations $\pi \in {\cal S}_n$. Due to (\ref{commute})
and the linearity of the action of the transfer matrix on $V^{\otimes n}$ it 
also commutes with $T_n$. The space
$U = Y\,V^{\otimes n} = \sum_{s^1,...,s^n} {\bf R} Y|s^1s^2...s^n\rangle$
is therefore invariant under action of $T_n$. If we are able to construct
elements of ${\bf s}_n$ projecting $V^{\otimes n}$ to a proper subspace 
we can thus achieve a block diagonalization of $T_n$ by its restriction to
$U$ and to its orthogonal complement $({\bf 1}-Y)\,V^{\otimes n}$.

\section{Some remarks on the symmetric group}

In this section we review some properties of the symmetric group and its
irreducible representations. These are well--studied and numerous 
excellent presentations can be found, e.g. in \cite{group1,group2}. Here
we omit any proofs.

The symmetric group ${\cal S}_n$ contains the $n!$ permutations of
$n$ distinguishable objects. Consider any representation $\tilde{D}$ on a
linear space $\tilde{V}$. $\tilde{D}$ is called to be irreducible 
iff there are no proper subspaces ($\neq \{0\}$) of $\tilde{V}$ 
closed under $\tilde{D}({\cal S}_n)$. A representation
is called completely reducible iff it can be decomposed into a
direct sum of irreducible subrepresentations. This decomposition is
unique up to isomorphisms. Our $D$ defined in the previous section
is completely reducible.

The irreducible representations of ${\cal S}_n$ are classified by
the so--called {\em standard Young tableaus}. Each Young tableau is 
characterized 
by a partition of $n$, i.e. a set of integers
 $\lambda_1\geq\lambda_2\geq ...\geq\lambda_m> 0,\; m\leq n$, fulfilling
$\sum_a\lambda_a=n$. One arranges m rows of length $\lambda_1,...,\lambda_m$
as shown in the figure and fills the boxes with the integers $1,...,n$. The
tableau is called standard iff the entries of the boxes are increasing within 
every row and within every column, see e.g. the figure.

\begin{figure}[htb]
 \epsfysize=4cm
      \epsffile{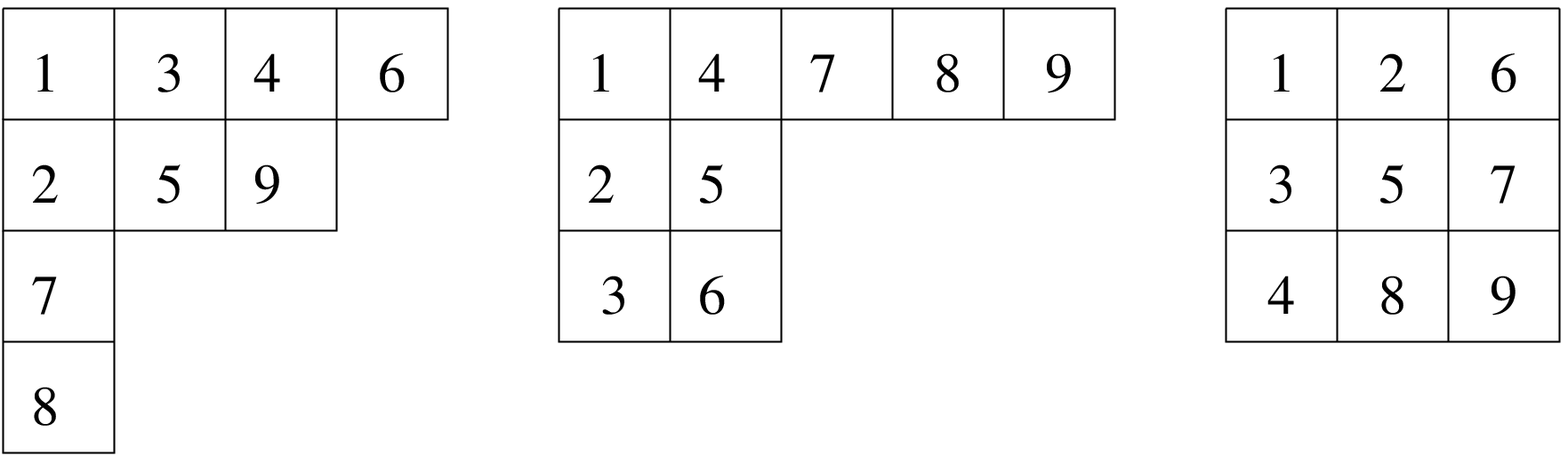}
\caption{Examples for standard Young tableaus for $n=9$}
\end{figure}

At first we define the row symmetrizer 
$\sym_{[\lambda_1,...,\lambda_m]} = \prod_{a=1}^{m} \sym_a$ with $\sym_a$ 
being the sum of all permutations within the $a$--th row. Then we still 
need the column antisymmetrizer 
$\as_{[\lambda_1,...,\lambda_m]} = \prod_{b=1}^{\lambda_1} \as_b$ with
$\as_b$ being the total antisymmetrizer of column $b$, i.e. the sum of
$(-1)^\pi \pi$ over all permutations of this column. $(-1)^\pi$ signifies
whether $\pi$ is odd or even.
The {\em Young operator} is then defined by 
\begin{equation}
  \label{Young}
  Y_{[\lambda_1,...,\lambda_m]} = 
  \as_{[\lambda_1,...,\lambda_m]}\sym_{[\lambda_1,...,\lambda_m]}
\end{equation}
and is an element of the group algebra ${\bf s}_n$. 

If we go back to the representation $\tilde{D}({\cal S}_n)$, 
then the action of
$Y_{[\lambda_1,...,\lambda_m]}$ on any element $|v\rangle$ of $\tilde{V}$ 
maps this vector to
an irreducible subrepresentation. A basis of the irreducible representation
space can be constructed by applying all permutations 
to $Y_{[\lambda_1,...,\lambda_m]}|v\rangle$
and selecting a maximal linearly independent subset. 
Every standard Young tableau gives a 
different irreducible representation, those corresponding to the same
partition $[\lambda_1,...,\lambda_m]$ but different entries are isomorphic.
Depending on the structure of $\tilde{D}$, also the action of the same Young
operator on different vectors from $\tilde{V}$ can give different irreducible
subrepresentations of $\tilde{D}$. Every irreducible subrepresentation can be 
constructed in the prescribed way.

Another notion needed in the following is that of the {\em associate}
representation. For any irreducible representation given by a standard
Young tableau with partition $[\lambda_1,...,\lambda_n]$ it is given by the
{\em transposed} standard Young tableau, i.e. the rows become the columns
and vice versa. The transposed partition is denoted by 
$[\tilde{\lambda}_1,...,\tilde{\lambda}_{\tilde m}]$ 
with $\tilde{\lambda}_1=m$
and $\tilde{m}=\lambda_1$. An example is given in the following figure.

\begin{figure}[htb]
 \epsfysize=4cm
      \epsffile{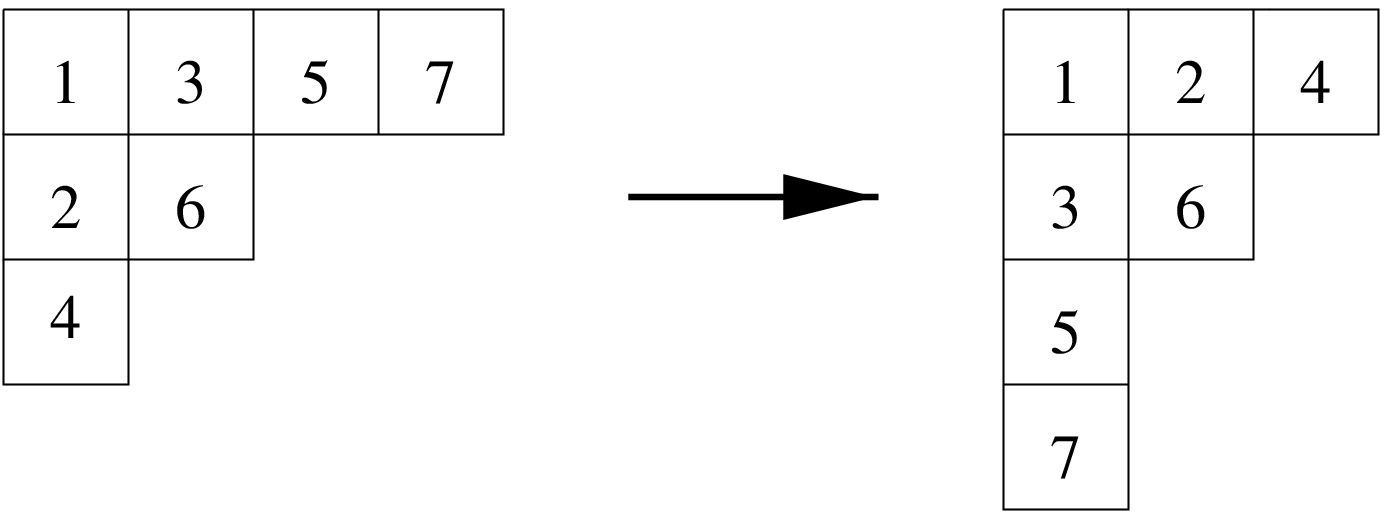}
\caption{Example for transposing a standard Young tableau}
\end{figure}

\section{The replica symmetric eigenspaces}

We now return to the problem of finding the eigenvalues of
the replicated and disorder--averaged transfer matrix $T_n$. At
the end of the second section we showed that the space $Y\,V^{\otimes n}$
is invariant with respect to $T_n$ for every $Y\in{\bf s}_n$.
In particular, this is the case for the Young operators which define
minimal invariant sets obtainable without further knowledge of 
the exact form of $T_n$.

In this section we concentrate on a special irreducible subrepresentation
described by the standard Young tableau with only one row. The Young
operator $Y_{[n]}$ becomes the symmetrizer of the complete symmetric group, 
its image $Y_{[n]}V^{\otimes n}$
is therefore invariant under permutations. The corresponding
irreducible subrepresentations of $D$ are thus one--dimensional,
all permutations are represented trivially by the identity. Consequently
the elements of $Y_{[n]}V^{\otimes n}$ are replica symmetric and therefore
also the eigenvectors of $T_n$ constructed within this space.

As basis vectors for $Y_{[n]}V^{\otimes n}$ we introduce
\begin{eqnarray}
  \label{rsvectors}
  |\rho_1,...,\rho_{R-1}\rangle& =& \frac{1}{\rho_1!\cdot...\cdot\rho_r!}
  Y_{[n]}\;|1\rangle^{\otimes\rho_1}\otimes...\otimes|R\rangle^{\otimes\rho_R}
  \nonumber\\
 & = & \sum_{\{s^a|\sum_a\delta_{s^a,s}=\rho_s\;
             \forall s=1,...,R-1\}}|s^1...s^n\rangle
\end{eqnarray}
where $\rho_s, s=1,...,R-1$, and $\rho_R=n-\sum_{s=1}^{R-1}\rho_s$ have to be
non--negative integers.
The replica symmetric submatrix of $T_n$ can be calculated by
\begin{equation}
  \label{rsmatrix}
  T_n^{[n]}(\rho_1,...,\rho_{R-1}|\sigma_1,...,\sigma_{R-1})
   = \frac{\langle\rho_1,...,\rho_{R-1}|T_n|\sigma_1,...,\sigma_{R-1}\rangle
    }{ \langle\rho_1,...,\rho_{R-1}|\rho_1,...,\rho_{R-1}\rangle}\;.
\end{equation}
The denominator results from the fact that the vectors (\ref{rsvectors}) are
orthogonal but not normalized. In order to send the replica number $n$ to
zero we have to introduce generating functions into the eigenvalue equation
\begin{equation}
  \label{eigenvals}
  \Lambda_{[n]} Z(\sigma_1,...,\sigma_{R-1}) = \sum_{\{\rho_1,...,\rho_{R-1}\}}
  T_n^{[n]}(\rho_1,...,\rho_{R-1}|\sigma_1,...,\sigma_{R-1})
  Z(\rho_1,...,\rho_{R-1})
\end{equation}
by writing
\begin{equation}
  \label{genfunc}
  \Phi[x_1,...,x_{R-1}] = \sum_{\{\rho_1,...,\rho_{R-1}\}}
  x_1^{\rho_1}\cdot...\cdot x_{R-1}^{\rho_{R-1}} Z(\rho_1,...,\rho_{R-1}).
\end{equation}
The eigenvalue equation (\ref{eigenvals}) now reads
\begin{equation}
  \label{laplace}
  \Lambda_{[n]} \Phi[x_1,...,x_{R-1}] 
   =\left( \sum_{s=1}^R x_s T_{R,s} \right)^n
    \cdot \Phi\left[ \frac{\sum_sx_sT_{1,s}}{\sum_sx_sT_{R,s}},...,
                   \frac{\sum_sx_sT_{R-1,s}}{\sum_sx_sT_{R,s}} \right]
\end{equation}
where we introduced $x_R=1$ for simplicity, for the calculations
 see appendix A.
In this equation a sensible limit $n\to 0$ can be performed. 
The largest eigenvalue is $\Lambda=1-\beta n f +O(n^2)$ 
where $\beta$ is the inverse temperature and $f$ the free energy.
Finally, we change from left to right eigenfunctions,
 introduce ${\boldx}=(x_1,...,x_{R-1})\in {\bf R}^{R-1}$ and
 \begin{equation}
   \label{f}
   h_r({\boldx}) = \frac{\sum_{s=1}^R x_s T_{r,s}}{\sum_{s=1}^R x_s T_{R,s}}
 \end{equation}
and obtain an equation for a $(R-1)$--dimensional invariant density
\begin{equation}
  \label{density}
  \Phi^{(0)}[{\boldx}] = \int d^{R-1}y \ll 
        \delta^{(R-1)}({\boldx}- {\boldh}({\boldy})) \gg \Phi^{(0)}[{\boldy}]
\end{equation}
where we used the $(R-1)$--dimensional Dirac distribution 
$\delta^{(R-1)}(\cdot)$. The density has to be normalized, 
$\int d^{(R-1)}x \; \Phi^{(0)}[{\boldx}] =1$  \cite{remark}.
As in perturbation theory we 
calculate the $O(n)$--corrections of $\lambda$ with the unperturbated
eigenfunction,
\begin{equation}
  \label{freeenergy}
  f = -\frac{1}{\beta} \int d^{(R-1)}x  \; \Phi^{(0)}[{\boldx}]
      \ll ln( \sum_{s=1}^{R} x_s T_{R,s} ) \gg \;.
\end{equation}
In this paper we do not calculate this free energy for any special 
distribution. 
This task itself is very hard and has been solved only for a few distributions
of quenched disorder, see e.g. \cite{Luck,Crisanti} and references therein.

The same equations can be obtained also without using replicas. For the
one--dimensional Ising model this was established by Derrida and Hilhorst in 
\cite{Derr}, their method using Riccati variables also generalizes to more 
complicated degrees of freedom than Ising spins. 
Another result reminiscent of ours was 
obtained by Lin \cite{Lin}, who showed the equivalence of an early replica
approach by Kac with Dyson's method for the phonon spectrum of a chain of
random masses and springs.

\section{Two--point correlations}

To be sure that replica symmetry is not violated we have to consider
the eigenvalues of the other representations. They have a very simple
interpretation in terms of connected two--point correlation functions.
  
Consider the operator
\begin{equation}
  \label{spinop}
  X |s\rangle = x_s |s\rangle
\end{equation}
where $x_s$ is any observable assigned to the basis vectors $|s\rangle$,
e.g. spin, location, or occupation number. 
It can be simply extended to the replicated
vector space $V^{\otimes n}$ by introducing the $n$ operators
$X_a^{(n)}={\bf 1}^{\otimes a-1}\otimes X\otimes{\bf 1}^{\otimes n-a}$,
$a=1,...,n$.
They are commutative and measure the value of $x$ at the $a$--th replica
site. In addition we introduce the operators
\begin{equation}
  X^{(\lambda)} = \left\{
    \begin{array}{ll}
    {\bf 1} & {\mbox{ if }}\lambda=1\\
    \prod_{1\leq a < b \leq \lambda} (X_a^{(\lambda)}-X_b^{(\lambda)}) &
         {\mbox{ if }}\lambda >1
    \end{array}
    \right.
\end{equation}
for any non--negative integer $\lambda$. For every partition 
$[\lambda_1,...,\lambda_m]$ and its transpose
$[\tilde{\lambda_1},...,\tilde{\lambda}_{\tilde{m}}]$,
they can be combined to the operator
\begin{equation}
 \label{Xop}
 X_{[\lambda_1,...,\lambda_m]}= 
 X^{(\tilde{\lambda_1})}\otimes...\otimes X^{(\tilde{\lambda}_{\tilde{m}})}
\end{equation}
acting on $V^{\otimes n}$.
Moreover, it maps any replica symmetric vector to a vector in a
representation space belonging to the standard Young tableau with $m$
rows of length
$\lambda_1,...,\lambda_m$ where we fill one column after the other
successively with integers $1,...,n$. An example is given by the 
2nd tableau in figure 1.
A sketch of the proof will be shown in appendix B.

Using this we find that
\begin{equation}
  \label{corr}
  {\mbox{tr}}( T_n^i X_{[\lambda_1,...,\lambda_m]} T_n^{j-i} 
  X_{[\lambda_1,...,\lambda_m]} T_n^{N-j}) \propto 
  \Lambda_{[\lambda_1,...,\lambda_m]}^{|j-i|}
\end{equation}
for large distances $|j-i|$. $\Lambda_{[\lambda_1,...,\lambda_m]}$
is the largest eigenvalue if we consider only eigenfunctions
in the subspace $Y_{[\lambda_1,...,\lambda_m]}\;V^{\otimes n}$.

Because of $\sum_a\lambda_a=n$ we have by definition (\ref{Xop})
\begin{equation}
  \label{Xop1}
  X_{[\lambda_1,...,\lambda_m]}=X_{[\sum_{a=2}^{m}\lambda_a,\lambda_2,
   ...,\lambda_m]} \otimes {\bf 1}^{n-\sum_{a=2}^{m}\lambda_a} \;.
\end{equation}
Introducing this into (\ref{corr}) we can send $n\to 0$ and obtain
\begin{equation}
  \label{corr1}
  \ll <X_{[\sum_{a=2}^{m}\lambda_a,\lambda_2,...,\lambda_m]}(i)\cdot
       X_{[\sum_{a=2}^{m}\lambda_a,\lambda_2,...,\lambda_m]}(j)> \gg
   \;\propto\; \lim_{n\to 0}
   \Lambda_{[n-\sum_{a=2}^{m}\lambda_a,\lambda_2,...,\lambda_m]}^{|j-i|}
   \;,
\end{equation}
i.e. the two-point correlation function of 
$X_{[\sum_{a=2}^{m}\lambda_a,\lambda_2,...,\lambda_m]}$
decays exponentially with correlation length 
$\xi=-1/\ln\Lambda_{[-\sum_{a=2}^{m}\lambda_a,\lambda_2,...,\lambda_m]}$.
$<\cdot>$ denotes the thermodynamic average in the disordered system
with transfer matrices $T_i, i=1,...,N$.
In order to calculate this we still need $2\;\sum_{a=2}^{m}\lambda_a$
real non--interacting replicas of the original quenched system.

Here we concentrate on Young tableaus
having only two rows, i.e. to partitions $[n-\lambda,\lambda]$. There
the operator reads
\begin{equation}
  \label{X2row}
  X_{[n-\lambda,\lambda]}= 
    (X\otimes{\bf 1}-{\bf 1}\otimes X)^{\otimes\lambda}
    \otimes {\bf 1}^{\otimes(n-2\lambda)}
\end{equation}
and consequently
\begin{equation}
  \label{conncorr}
  \ll <X_{[\lambda,\lambda]}(i)\cdot X_{[\lambda,\lambda]}(j)> \gg
  \;=\; \ll (<x_{s_i}x_{s_j}> - <x_{s_i}><x_{s_j}>)^\lambda \gg
  \propto \Lambda_{[-\lambda,\lambda]}^{|j-i|}
\end{equation}
describes the $\lambda$--th moment of the connected two--point
correlation function with respect to the disorder distribution.
The correlation length diverges whenever 
$\lim_{n\to 0}\Lambda_{[n-\lambda,\lambda]} = 1$. The criterion for
replica symmetry breaking, i.e. the degeneracy of the largest
replica symmetric eigenvalue with a non--symmetric one, thus
coincides with the standard criterion for a phase transition.
However, for one--dimensional systems with finite $R$ we cannot
expect any phase transition for non--vanishing temperature.

In App. C we will develop an equation for
$\lim_{n\to 0}\Lambda_{[n-\lambda,\lambda]}$. The calculations
are quite similar to replica symmetric one, but due to the more
complicated representation structure they are somewhat lengthy.
Here we give only the final result, an eigenvalue equation for
a function
\begin{equation}
  \label{Phi}
  \boldphi^{[-\lambda,\lambda]} : {\bf R}^{R-1} \to {\bf R}^{((R-1)^\lambda)}
\end{equation}
given by its components 
$\Phi_{s^1,...,s^\lambda}^{[-\lambda,\lambda]}(\boldx)$:
\begin{equation}
  \label{correv}
  \Lambda_{[-\lambda,\lambda]}
    \Phi^{[-\lambda,\lambda]}_{s^1,...,s^\lambda}(\boldx) =
  \int d^{R-1}y \; \sum_{r^1,...,r^\lambda=1}^{R-1}
   \; \ll \delta^{(R-1)}(\boldx -\boldh(\boldy))
  \prod_{a=1}^{\lambda}
    \frac{\partial h_{s^a}}{\partial y_{r^a}} \gg \;
    \Phi_{r^1,...,r^\lambda}^{[-\lambda,\lambda]}(\boldx)
\end{equation}
For every eigenfunction $\boldphi^{[-1,1]}(\boldx)$ of 
$T_n^{[n-1,1]}$ for $n\to 0$  the function
$\nabla\cdot\boldphi^{[-1,1]}(\boldx)$ is an eigenfunction
of the replica symmetric transfer matrix given in (\ref{density})
to the same eigenvalue.
Only the largest replica symmetric eigenvalue (=1) cannot be
reached in this way, because the integral of 
$\nabla\cdot\boldphi^{[-1,1]}(\boldx)$ over the definition space
${\bf R}^{R-1}$ vanishes. Therefore the largest eigenvalue of
$T_0^{[-1,1]}$ equals the second largest of $T_0^{[0]}$
and so on. The first transfer matrix block which could
produce a diverging correlation length outside the replica symmetric
sector is the one corresponding to $[-2,2]$, i.e. to the second
moment of the connected two--point correlation function. This is know
to be just for spin glass transitions where the second moment of the
connected two--point function describes 
the non--linear susceptibility, cf. \cite{binder}.

\section{Summary and outlook}

In this paper we developed a general replica transfer matrix method
capable of handling products of random finite--dimensional matrices.
We obtained expressions for the free energy (or Lyapunov exponent)
from the replica symmetric eigenvalues
and for the correlation length of the moments of the connected
correlation function from non--trivial representations of
the symmetric group. So we showed that the representation theoretic
approach to replica symmetry breaking is a general tool for
one--dimensional models and probably it can be extended to
two--dimensional models by considering larger and larger
one--dimensional stripes. We have
not applied our results to particular models, this should
done in some future work in order to study the possibility
of replica symmetry breaking in one-- or two--dimensional
systems.

{\bf Acknowledgment}: I am very grateful to J. Berg and R. Monasson for
fruitful discussions and careful reading the manuscript.

\appendix 

\section{Laplace transform of the replica symmetric eigenvalue equation}

In this appendix we calculate the Laplace transform of the
replica symmetric eigenvalue equation (\ref{eigenvals}). We start
with
\begin{equation}
  \label{Aeigenvals}
  \Lambda_{[n]} Z(\sigma_1,...,\sigma_{R-1}) = \sum_{\{\rho_1,...,\rho_{R-1}\}}
  T_n^{[n]}(\rho_1,...,\rho_{R-1}|\sigma_1,...,\sigma_{R-1})
  Z(\rho_1,...,\rho_{R-1})
\end{equation}
where the replica symmetric transfer matrix is given by
\begin{equation}
  \label{Arsmatrix}
  T_n^{[n]}(\rho_1,...,\rho_{R-1}|\sigma_1,...,\sigma_{R-1})
   = \frac{\langle\rho_1,...,\rho_{R-1}|T_n|\sigma_1,...,\sigma_{R-1}\rangle
    }{ \langle\rho_1,...,\rho_{R-1}|\rho_1,...,\rho_{R-1}\rangle}\;.
\end{equation}
using the replica symmetric vectors
\begin{equation}
  \label{Arsvectors}
  |\rho_1,...,\rho_{R-1}\rangle =\sum_{\{s^a|\sum_a\delta_{s^a,s}=\rho_s\;
             \forall s=1,...,R-1\}}|s^1...s^n\rangle\;,
\end{equation}
see Sec. IV.
If we introduce the Laplace transformation (\ref{genfunc}) on the left side 
of (\ref{Aeigenvals}) we obtain 
(introducing $x_R=1, \rho_R=n-\rho_1-...-\rho_{R-1}$)
\begin{eqnarray}
  \label{Alaplace}
  \Lambda_{[n]} \Phi[x_1,...,x_{R-1}] &=& \sum_{\rho_1,...,\rho_{R-1}}
       Z(\rho_1,...,\rho_{R-1}) \sum_{\sigma_1,...,\sigma_{R-1}}
       x_1^{\sigma_1}\cdot...\cdot x_{R-1}^{\sigma_{R-1}}
       \;\;T_n^{[n]}(\rho_1,...,\rho_{R-1}|\sigma_1,...,\sigma_{R-1})
       \nonumber\\
  &=& \sum_{\rho_1,...,\rho_{R-1}} Z(\rho_1,...,\rho_{R-1})
      \;\;\sum_{\{s^a\}} x_1^{\Sigma_a\delta_{s^a,1}}\cdot...\cdot
      x_{R-1}^{\Sigma_a\delta_{s^a,R-1}}
      \;\;\langle1|^{\otimes\rho_1}\otimes...\otimes\langle R|^{\otimes\rho_R}
      T_n |s^1...s^n\rangle \nonumber\\
  &=& \sum_{\rho_1,...,\rho_{R-1}} Z(\rho_1,...,\rho_{R-1})
      \left( \sum_{s=1}^R x_s T_{1,s} \right)^{\rho_1}\cdot...\cdot
      \left( \sum_{s=1}^R x_s T_{R,s} \right)^{\rho_R}\nonumber\\
  &=& \left( \sum_{s=1}^R x_s T_{R,s} \right)^{n} 
      \sum_{\rho_1,...,\rho_{R-1}} Z(\rho_1,...,\rho_{R-1})
      \prod_{r=1}^{R-1} \left( 
        \frac{\sum_{s=1}^R x_s T_{r,s}}{\sum_{s=1}^R x_s T_{R,s}}
                        \right)^{\rho_r}  \nonumber\\
  & =&\left( \sum_{s=1}^R x_s T_{R,s} \right)^n
  \cdot \Phi\left[ \frac{\sum_sx_sT_{1,s}}{\sum_sx_sT_{R,s}},...,
                   \frac{\sum_sx_sT_{R-1,s}}{\sum_sx_sT_{R,s}} \right]
\end{eqnarray}
which is equation (\ref{laplace}).

\section{Proof to Sec. V}

In this appendix we show that the operators 
$X_{[\lambda_1,...,\lambda_m]}$ defined in (\ref{Xop}) map any 
replica symmetric
vector to a representation space for an irreducible representation with a
Young tableau described by $[\lambda_1,...,\lambda_m]$.
This can be done by proving the equation
\begin{equation}
  \label{rsmaptorsb}
  Y_{[\lambda_1,...,\lambda_m]}\;X_{[\lambda_1,...,\lambda_m]}\;Y_{[n]}
 = c_{[\lambda_1,...,\lambda_m]}\;X_{[\lambda_1,...,\lambda_m]}\;Y_{[n]}
\end{equation}
where $c_{[\lambda_1,...,\lambda_m]}$ is a real number given by
$Y_{[\lambda_1,...,\lambda_m]}^2$ =
$c_{[\lambda_1,...,\lambda_m]} Y_{[\lambda_1,...,\lambda_m]}$.
Here we concentrate on the case $[n-\lambda,\lambda]$, i.e. to Young
tableaus with only two rows. These are the most important cases for
our needs, and the proof can be generalized directly to more
complicated tableaus as well using analogous procedures.

In the case of two rows we have
\begin{equation}
  \label{Ylambda}
  Y_{[n-\lambda,\lambda]}= (1-(1,2))(1-(3,4))...(1-(2\lambda-1,2\lambda))
  \cdot \sym_{[n-\lambda,\lambda]}
\end{equation}
where $(a,b)$ denotes the transposition permuting $a$ and $b$, and
\begin{equation}
  \label{Xlambda}
  X_{[n-\lambda,\lambda]}=(X\otimes{\bf 1}-{\bf 1}\otimes X)^{\otimes\lambda}
    \otimes{\bf 1}^{\otimes(n-2\lambda)}
\end{equation}

(i) As a first step we note that
\begin{eqnarray}
  \label{proof1}
  \forall \pi \in{\cal S}_n: \;\;\;
  \pi\; X_{a_1}^{(n)}\cdot...\cdot X_{a_l}^{(n)}\;Y_{[n]}& = &
   X_{\pi(a_1)}^{(n)}\cdot...\cdot X_{\pi(a_l)}^{(n)}\;\pi\;Y_{[n]}
\nonumber\\ &=&X_{\pi(a_1)}^{(n)}\cdot...\cdot X_{\pi(a_l)}^{(n)}\;Y_{[n]}\;.
\end{eqnarray}
It follows that $\sym_{[n-\lambda,\lambda]} X_{[n-\lambda,\lambda]}Y_{[n]}$
is a sum of certain $X_{a_1}^{(n)}\cdot...\cdot X_{a_\lambda}^{(n)}\;Y_{[n]}$
with integer prefactors depending on $a_1 < ... < a_\lambda$.

(ii) The action of 
$\as_{[n-\lambda,\lambda]}= (1-(1,2))(1-(3,4))...(1-(2\lambda-1,2\lambda))$
on these gives
\begin{equation}
  \label{proof2}
  \as_{[n-\lambda,\lambda]}\;
  X_{a_1}^{(n)}\cdot...\cdot X_{a_\lambda}^{(n)}\;Y_{[n]} = \left\{
    \begin{array}{lll}
      \pm X_{[n-\lambda,\lambda]}\;Y_{[n]}\;\;\;&\mbox{if}&
         a_\rho\in\{2\rho-1,2\rho\}\;\;\forall \rho =1,...,\lambda\\
    0 & \mbox{else}&
    \end{array}
    \right.
\end{equation}
If there were the factors $X_{2\rho-1}^{(n)}$ and $X_{2\rho}^{(n)}$
for any $\rho\leq\lambda$, the action of $(1-(2\rho-1,2\rho))$ would
annihilate the term. The same happens, if there is any $\rho\leq\lambda$
for which neither $X_{2\rho-1}^{(n)}$ nor $X_{2\rho}^{(n)}$ appear
in the product. The sign in (\ref{proof2}) can be obtained by counting
the even indices $a_\rho$ in $X_{a_1}^{(n)}\cdot...\cdot X_{a_\lambda}^{(n)}$.

Altogether we find that the action of the Young operator produces only
a constant of proportionality, and the proof is complete.

\section{Calculation of non--symmetric eigenvalue equations}

In this appendix we present the calculation of the eigenvalue
equations for non--trivial irreducible representations at the
example of $[n-1,1]$. This case is surely the simplest non--trivial
one, but the ideas of the calculation are the same also for
higher representations.

We consider the standard Young tableau for the partition  $[n-1,1]$
having entries $1,3,4,...,n$ in the first row and $2$ in the
second. The corresponding Young operator
\begin{equation}
  \label{Ayoung}
  Y_{[n-1,1]} = (1-(1,2))\cdot \sym(1,3,4,...,n)
\end{equation}
maps an arbitrary basis vector $|s^1...s^n\rangle$ 
up to a normalization constant to
\begin{equation}
  \label{D1vecs}
  |s^2;\sigma_1,...,\sigma_{R-1}\rangle := \sum_{s\neq s^2}
   (|ss^2\rangle-|s^2s\rangle)\otimes
   |\sigma_1,...,\sigma_{\mbox{min}(s,s^2)}-1,...,\sigma_{\mbox{max}(s,s^2)}-1,
   ...,\sigma_{R-1}\rangle
\end{equation}
where the last term in the product is a symmetrized vector in the
$(n-2)$--fold replicated vector space, cf. equation (\ref{rsvectors}), and
$\sigma_s=\sum_{a=1}^n \delta_{s^a,s}$. Due to
$Y_{[n-1,1]} T_n = T_n Y_{[n-1,1]}$ these vectors form an invariant set
with respect to $T_n$. For given $\sigma_1,...,\sigma_{R-1}$ there are
$R-1$ linearly independent vectors of this type, so without loss of
generality we can choose $s^2=1,...,R-1$.

The transfer matrix block to be calculated is
\begin{eqnarray}
  \label{transfer}
  T_n^{[n-1,1]}(s;\sigma_1,...,\sigma_{R-1}|r;\rho_1,...,\rho_{R-1})& =&
  \frac{\langle s;\sigma_1,...,\sigma_{R-1}|T_n|r;\rho_1,...,\rho_{R-1}
    \rangle}{\langle s;\sigma_1,...,\sigma_{R-1}| s;\sigma_1,...,\sigma_{R-1}
        \rangle}\nonumber\\
  &=& \sum_{t\neq r} (T_{R,t}T_{s,r}-T_{R,r}T_{s,t})\nonumber\\
  && \;\;\;\;\;\;\times  (T^{\otimes(n-2)})^{[n-2]}
  (\sigma_1,...,\sigma_s-1,...,\sigma_{R-1}|\rho_1,...,\rho_r-1,...,\rho_{R-1})
  \;.
\end{eqnarray}
The matrix $(T^{\otimes(n-2)})^{[n-2]}$ is nothing but the replica
symmetric matrix $T_{n-2}^{[n-2]}$ {\it without} the average over the
quenched disorder. For the eigenvalue equation
\begin{equation}
  \label{D1eigen}
  \Lambda_{[n-1,1]} \cdot C(r;\rho_1,...,\rho_{R-1})=
  \sum_{s;\sigma_1,...,\sigma_{R-1}} 
  T_n^{[n-1,1]}(s;\sigma_1,...,\sigma_{R-1}|r;\rho_1,...,\rho_{R-1})
  C(s;\sigma_1,...,\sigma_{R-1})
\end{equation}
we introduce again a Laplace transform by
\begin{equation}
  \label{D1laplace}
  \Phi_s[x_1,...,x_{R-1}] = \sum_{\sigma_1,...,\sigma_{R-1}}
  x_1^{\sigma_1}\cdot...\cdot x_s^{\sigma_s-1}\cdot...\cdot
  x_{R-1}^{\sigma_{R-1}}\; C(s;\sigma_1,...,\sigma_{R-1}) \;.
\end{equation}
Due to ($x_R:=1$)
\begin{eqnarray}
&&  \sum_{\rho_1,...,\rho_{R-1}} x_1^{\rho_1}\cdot...\cdot 
  x_r^{\rho_r-1}\cdot...\cdot x_{R-1}^{\rho_{R-1}}\;
  T_n^{[n-1,1]}(s;\sigma_1,...,\sigma_{R-1}|r;\rho_1,...,\rho_{R-1})\nonumber\\
& =&\ll \sum_{t\neq r}(T_{R,t}T_{s,r}-T_{R,r}T_{s,t})\;x_t\left(
  \sum_{\rho_1,...,\rho_{R-1}}  x_1^{\rho_1}\cdot...\cdot x_r^{\rho_r-1}
  \cdot...\cdot x_t^{\rho_t-1}\cdot...\cdot x_{R-1}^{\rho_{R-1}}
  \; (T^{\otimes(n-2)})^{[n-2]}(...|...)\right)\gg\nonumber\\
& = &\ll \sum_{t\neq r}(T_{R,t}T_{s,r}-T_{R,r}T_{s,t})\;x_t
  \left(\sum_{p=1}^{R} x_p T_{R,p}\right)^{n-2} 
  \prod_{q=1}^{R-1}\left(
  \frac{\sum_{p=1}^{R} x_p T_{q,p}}{\sum_{p=1}^{R} x_p T_{R,p}}
    \right)^{\sigma_q-\delta_{q,s}}\gg \nonumber\\
& = & \ll \left(\sum_{p=1}^{R} x_p T_{R,p}\right)^n \cdot
   \frac{\partial h_s(\boldx)}{\partial x_r}\cdot
   \prod_{q=1}^{R-1} h_q(\boldx)^{\sigma_q-\delta_{q,s}}\gg\;,
\end{eqnarray}
where the 2nd last step is the same as in App. A for the replica symmetric
case, and where we use the function $\boldh(\boldx)$ defined in (\ref{f}),
the eigenvalue equation becomes
\begin{equation}
  \label{D1evlapl}
  \Lambda_{[n-1,1]}\;\Phi_r[x_1,...,x_{R-1}] =
  \left(\sum_{p=1}^{R} x_p T_{R,p}\right)^n \sum_{s=1}^{R-1}
  \frac{\partial h_s}{\partial x_r}\;\Phi_s[h_1(\boldx),...,h_{R-1}(\boldx)]\;.
\end{equation}
In the limit $n\to 0$ this results in equation (\ref{correv}) for
$\lambda=1$. The calculations for larger $\lambda$ are analogous.


\begin{thebibliography}{[99]}
\bibitem{Luck} J.M. Luck, {\it Syst\`emes d\'esordonn\'es unidimensionels} 
(Al\'ea Saclay, 1992)
\bibitem{Crisanti} A. Crisanti, G. Paladin, and A. Vulpiani,
{\it Products of Random Matrices in Statistical Physics} (Springer--Verlag
Berlin Heidelberg, 1993)
\bibitem{Pendry} J. B. Pendry, {\it Adv. Phys.} {\bf 43}, 461 (1994)
\bibitem{MPV} M. M\'ezard, G. Parisi, and M. A. Virasoro, {\it Spin
Glass Theory and Beyond} (World Scientific Singapore, 1987)
\bibitem{Newman} C. M. Newman and D. L. Stein, preprint cond-mat/9612097
\bibitem{RFIM1D} M. Weigt and R. Monasson, {\it Europhys. Lett.} {\bf 36},
209 (1996)
\bibitem{group1} W. Ludwig and C. Falter, {\it Symmetries in Physics}
(Springer--Verlag Berlin Heidelberg, 1988)
\bibitem{group2} M. Hamermesh,{\it Group Theory and its Application to
Physical Problems} (Addison Wesley, Reading, Mass. 1964)
\bibitem{remark} This last point still remains somewhat mysterious -- 
by changing from left to right eigenfunctions we also change the function 
space from polynomials to functions having a finite integral. Up to now
this step is only justified by its results (\ref{density},\ref{freeenergy})
and there coincidence with the results of \cite{Derr}.
\bibitem{Derr} B. Derrida, H. J. Hilhorst, {\it J. Phys. A}
{\bf 16}, 2641 (1983) 
\bibitem{Lin} T.F. Lin, {\it J. Math. Phys.} {\bf 11}, 1584 (1970)
\bibitem{binder} K. Binder and A. P. Young, {\it Rev. Mod. Phys.}
{\bf 58}, 801 (1986)
\end{thebibliography}
\end{document}